\documentclass[submission,copyright,creativecommons,noderivs,noncommercial]{eptcs}
\providecommand{\event}{FESCA 2014} 
\usepackage{breakurl}             
\usepackage[pdftex]{graphicx}
\usepackage{caption}
\usepackage{subcaption}
\usepackage{xspace}

\newcommand{\eg}{e.g.,\xspace}
\newcommand{\ie}{i.e.,\xspace}

\title{Execution Time Analysis for Industrial Control Applications}
\author{Stefan Stattelmann
\institute{ABB Corporate Research\\
Ladenburg, Germany}
\email{stefan.stattelmann@de.abb.com}
\and
Manuel Oriol
\institute{ABB Corporate Research\\
Baden-D\"attwil, Switzerland}
\email{manuel.oriol@ch.abb.com}
\and
Thomas Gamer
\institute{ABB Corporate Research\\
Ladenburg, Germany}
\email{thomas.gamer@de.abb.com}
}
\def\titlerunning{Timing Analysis for Industrial Control}
\def\authorrunning{S. Stattelmann, M. Oriol \& T. Gamer}
\begin{document}
\maketitle

\begin{abstract}
Estimating the execution time of software components is often mandatory when
evaluating the non-functional properties of software-intensive systems.  This
particularly holds for real-time embedded systems, e.g., in the context of
industrial automation.  In practice it is however very hard to obtain reliable
execution time estimates which are accurate, but not overly pessimistic with
respect to the typical behavior of the software.


This article proposes two new concepts to ease the use of execution time
analysis for industrial control applications: (1) a method based on recurring occurrences of code sequences for automatically creating a timing model of a given processor and (2) an interactive way to integrate execution time analysis into the development environment, thus making timing analysis results easily accessible for software developers.
The proposed methods are validated by an industrial case study, which 
shows that a significant amount of code reuse is present in a set of representative
industrial control applications.
\end{abstract}

\section{Introduction}\label{sec:introduction}
Calculating the execution time for a given piece of code on a modern processor
is a very difficult problem. In most cases, an accurate solution is impossible to find because it relates to the halting problem which is
undecidable.  In real-time systems, however, calculating the longest time a
piece of code might take to execute (Worst-Case Execution time, WCET) is necessary to verify that all real-time requirements are met.  To obtain
an estimate for the WCET despite its general undecidability, approximation
techniques must be used. There are two main approaches to approximate the WCET
of a software component: Testing or in-situation profiling of a piece of code
can be used to measure the execution time of diverse executions. The WCET of
the respective code can then be estimated from these measurements through
heuristics.  Alternatively, static code analysis techniques can be used to
calculate the WCET using a model of the hardware.

The first approach requires that sufficient code coverage can be achieved
during the measurements and that the worst-case execution time for every
program statement has been observed.  In practice, these requirements cannot
always be fulfilled and therefore the WCET might be under-approximated.
The second approach, static WCET analysis, requires the manual development of a
hardware model for the target processor, which estimates the execution time
based on formal methods. This is a long and costly process, as it requires the
efforts of highly specialized software developers for multiple person months.
The abstraction underlying the model approximates the execution time in a
conservative fashion, thereby leading to an over-approximation of the actual
WCET.  On the positive side, the resulting WCET estimate can be safely applied
in systems with hard real-time requirements, as it is guaranteed to be an upper
bound for every possible execution time of the program.

Besides the manual development of the timing model for the hardware, the
integration of the execution time estimates into the development process of
software with real-time constraints is another challenge. The time a piece of
software requires to execute on a specific processor can be expressed with one
or more of the following three values: best-case execution time (BCET),
worst-case execution time (WCET) and average-case execution time (ACET).
Existing static analysis tools only approximate the best-case and worst-case
execution time, usually by determining an upper or lower bound for the possible
execution times of a program. As the interaction and interdependence between
program parts is only approximated, the resulting estimates often contain
execution paths through the program which are infeasible in practice. The abstraction underlying these estimates can both hide the semantics of the programs and the execution time estimates can deviate significantly from the
average execution time observable during measurements.

The problem with determining only an upper and lower bound for the execution
time is that this information is of little use for software developers of soft
real-time or non-real-time systems.  For this usage scenario, the average-case
execution time is more relevant. Today, a large number of program runs has to
be observed and measured in order to determine an accurate average-case estimate or a
good approximation of the execution time distribution from measurements. Such
average-case execution time estimates can only be determined if hardware with
measurement or profiling capabilities is available \emph{during software
development}. 
There is however no established method
and no established tool for reasoning about average-case execution time without
observing actual program executions.

This paper proposes two new concepts to facilitate the use of execution time
analysis during software development for real-time applications. The techniques
were originally developed for industrial control applications, but are equally
applicable in other domains with (soft) real-time requirements.  
Section~\ref{sec:rw} introduces related work on worst-case execution time analysis.
Section~\ref{sec:timing} of this paper proposes a method to automatically create a timing
model for a given processor based on the repeated occurrence of
code patterns resulting from model-based code generation.  In Section~\ref{sec:user},
an interactive way to integrate execution time analysis into
the development environment is described, thereby improving the interaction
between timing analysis tools and software developers. In Section~\ref{sec:results},
preliminary results validating the applicability of the proposed approach are
presented. Finally, Section~\ref{sec:conc} summarizes the paper and gives an outlook.

\section{State of the Art in Execution Time Analysis}\label{sec:rw}
As mentioned in the previous section, execution time estimation techniques 
can be classified according to the way
the estimates are determined. Static techniques use a structural analysis of a
piece of software and an analytical model of the underlying hardware to
determine execution time estimates without executing the software. On the other
hand, dynamic techniques require executing the program of interest in order to
estimate the execution time of a program.  Furthermore, some dynamic estimation
techniques also use a (static) structural analysis of a program when estimating
its execution time.
Both types of techniques are explained in detail in the following paragraphs.

\subsection{Static Timing Analysis}
\label{sa_timing}

\begin{figure}
\centering
\begin{minipage}{.5\textwidth}
  \centering
  \includegraphics[height=.75\textwidth]{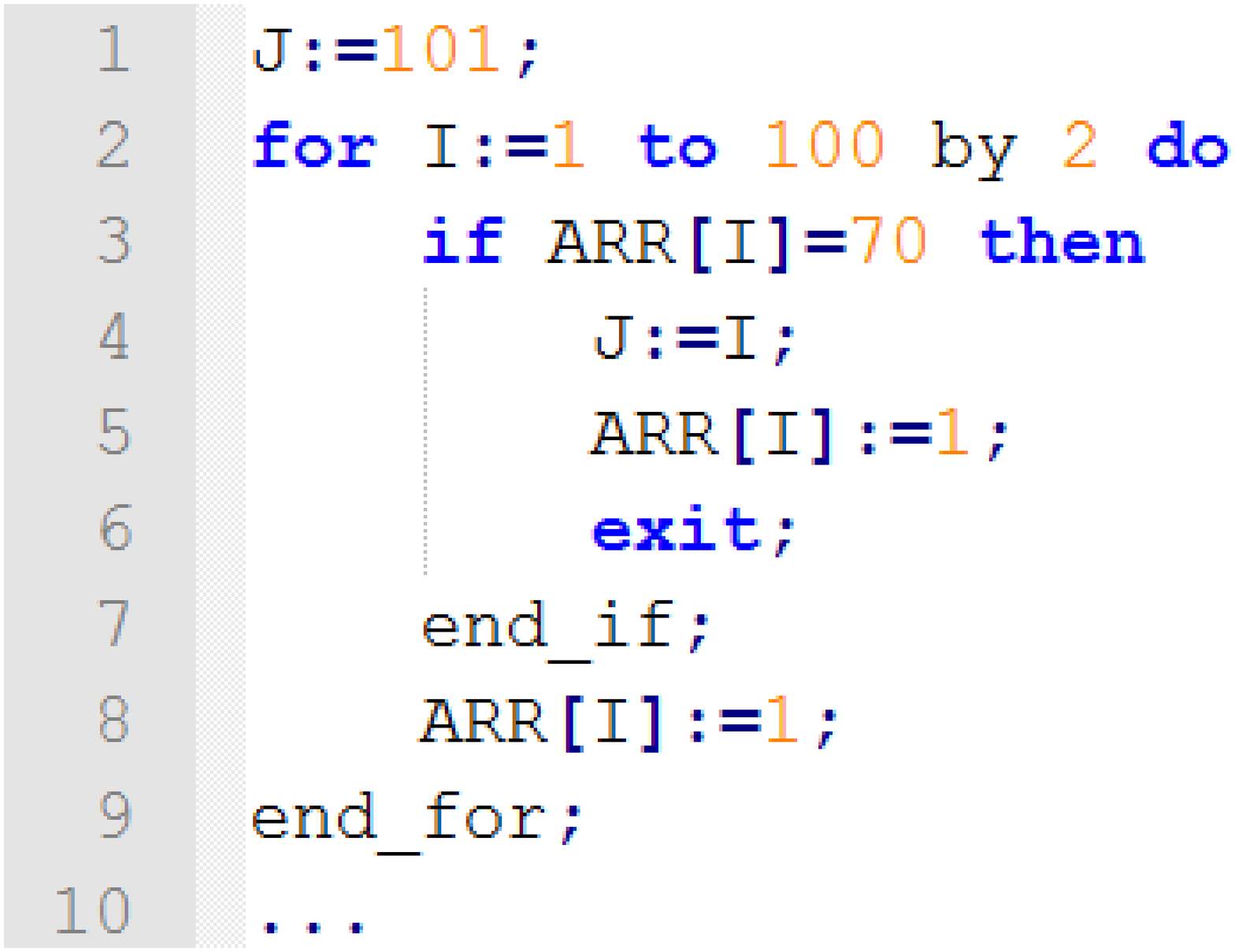}
  \captionof{figure}{Example of code}
  \label{fig:sample_code}
\end{minipage}%
\begin{minipage}{.5\textwidth}
  \centering
  \includegraphics[height=.75\textwidth, page=1]{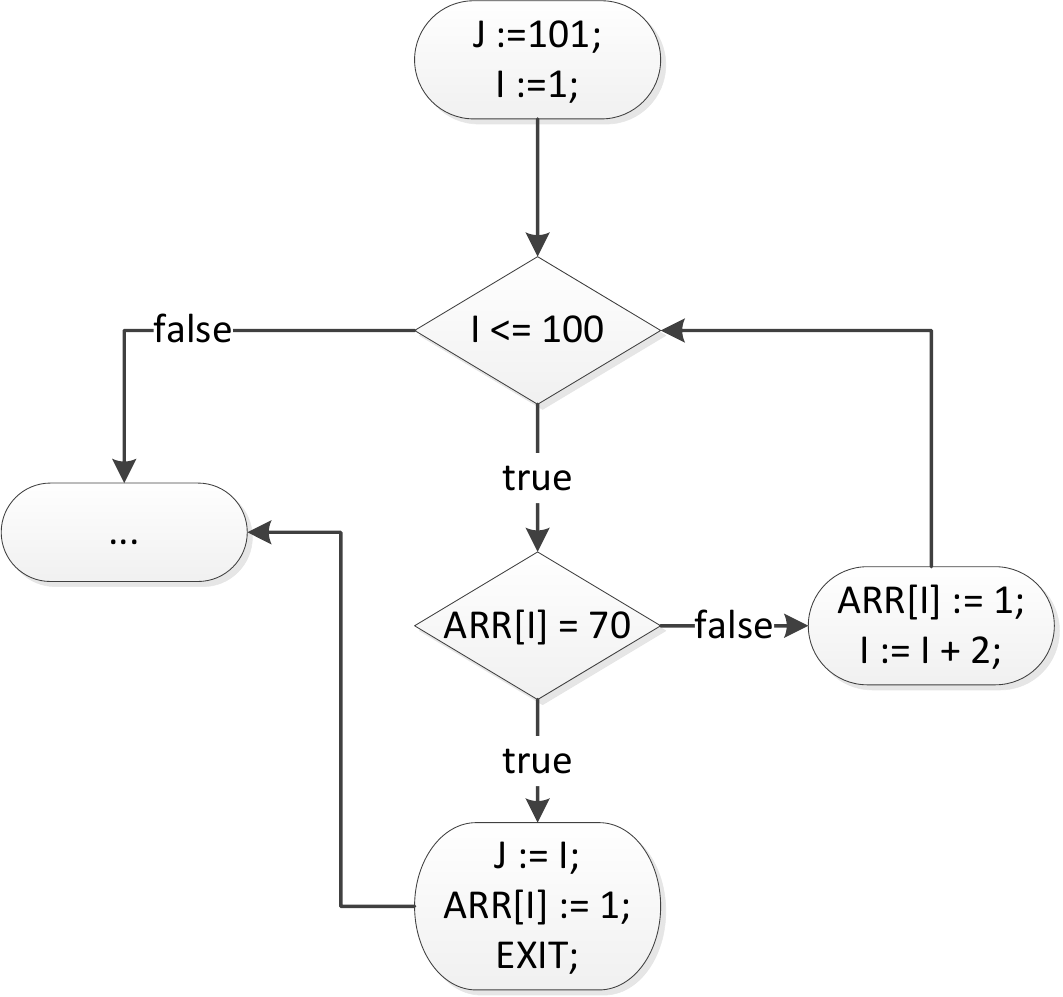}
  \captionof{figure}{Extracted control flow}
  \label{fig:CFG1}
\end{minipage}
\end{figure}

\begin{figure}
\centering
\begin{minipage}{.5\textwidth}
  \centering
  \includegraphics[height=.7\textwidth, page=2]{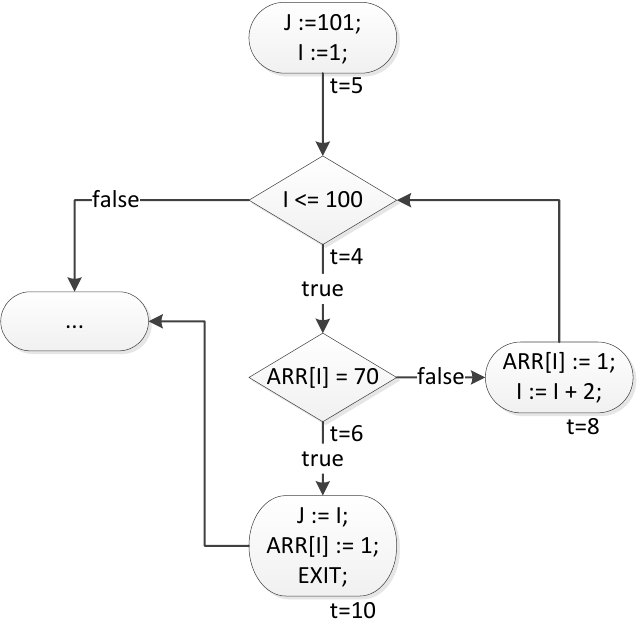}
  \captionof{figure}{Result of Micro-architectural Analysis}
  \label{fig:CFG2}
\end{minipage}%
\begin{minipage}{.5\textwidth}
  \centering
  \includegraphics[height=.7\textwidth, page=3]{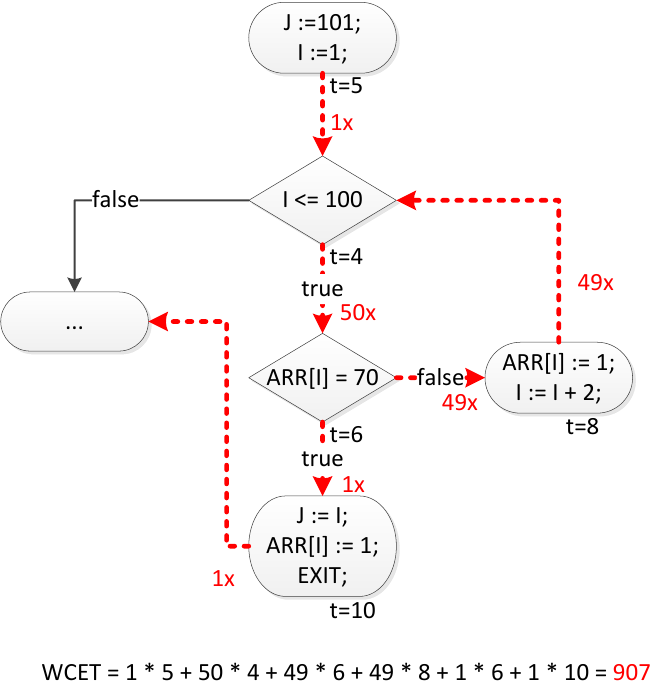}
  \captionof{figure}{Result of Global Bound Calculation}
  \label{fig:CFG3}
\end{minipage}
\end{figure}

Static timing analysis techniques estimate the execution time of a program
without actually executing any code. They are mainly used to determine 
the WCET of a program, meaning a conservative estimate or upper bound for
the execution time of a program. Most existing solutions are based on static
program analysis techniques to model the execution of a piece of software on a
given target processor~\cite{Wilhelm2008}. They can be roughly decomposed into the following three
steps (see
Figure~\ref{fig:sample_code} to Figure~\ref{fig:CFG3}):
\begin{enumerate}
\item \emph{Control flow analysis} decomposes the structure of the program into atomic
units for the subsequent analysis steps. The resulting program representation
is usually the control flow graph (CFG) of the program, which consist of basic
blocks.  A basic block is a maximal sequence of program statements with only
one point of entry and exit.  For the example in
Figure~\ref{fig:sample_code}, the resulting CFG is shown in
Figure~\ref{fig:CFG1}.
\item \emph{Micro-architectural analysis} determines the execution time for the atomic
units of a program using the result of the control flow analysis. In most cases
this analysis is performed using an abstract model of the target processor.
This model can be based on abstract interpretation~\cite{Theiling2000} or
symbolic execution~\cite{Lundqvist99}. In both cases the model focuses on the
execution time for sequences of machine instructions and can neglect details of
the computations these instructions would perform on the real hardware. This
abstraction is necessary to make the estimations computationally feasible, but
it can also be the source of imprecision.  The reason for this is the way the
micro-architectural analysis deals with information about the system state
which is not available due to abstraction: whenever the analysis cannot prove
that certain effects which impact the execution time, e.g., cache misses, will
never occur at a certain program point, it will assume the system state leading
to the longest execution time.  Depending on how much abstraction is necessary
to cope with complexity of the internal state of the processor, this can entail
a large number of pessimistic assumptions and thus a large overestimation of
the WCET.  An example for the outcome of this analysis step is shown  in
Figure~\ref{fig:CFG2} by annotating execution times to the CFG from
Figure~\ref{fig:CFG1}. Time units are omitted for this example, but most
existing tools use CPU clock cycles.
\item \emph{Global bound calculation} uses the results of the two previous steps to
obtain an estimate for the total execution time of a program. The prevalent
technique for doing this is implicit path enumeration~\cite{Li1997}.  This
approach translates the structural constraints and local execution time
estimates into an integer linear programming (ILP) problem, which is then
solved using standard ILP solvers.  If a program contains loops, the maximal
number of times a loop may execute must be determined by a previous analysis or
provided by the user.  This is necessary for the ILP solver to find a
worst-case path. The loop in the example from Figure~\ref{fig:sample_code} can
be executed at most 50 times. Thus, the edge leaving the basic block checking
the loop condition can be taken at most 50 times on any execution path of the
program.  With this structural information and the cost associated with each
basic block, the global bound analysis will identify the worst-case path, as shown
using dotted lines in Figure~\ref{fig:CFG3}.  This path assumes that the maximal
number of loop iterations is executed and that the conditional statement in
line~3 of Figure~\ref{fig:sample_code}, which aborts the loop, will be executed
in the 50th iteration of the loop.
\end{enumerate}

Existing static timing analysis tools are able to estimate the
best-case or worst-case execution time of software components. Estimating an
average-case execution time is usually not supported. Moreover, the challenge with static
timing analysis is the creation of the required abstract processor models.
Currently, these models are developed manually by the tool vendors, which is a
tedious, costly and error-prone process.
For simple processor architectures, static WCET analysis can estimate execution
times with an error rate of less than 10\%. However, this high level of
accuracy can only be achieved if the program path predicted by the WCET
analysis roughly matches the execution path observed during the measurement
runs~\cite{Zhang2005}. For simple processor architectures with simple pipelines
and without caches, the third step, global bound calculation, is therefore the most
likely phase to introduce an overestimation. On the other hand, for processors
with multi-level caches and multiple cores, the micro-architectural analysis
can also introduce a lot of pessimism to the execution time estimates as
internal state of the processor must be approximated conservatively.
Consequently, the obtainable accuracy of the execution time estimates reported
by static WCET analysis tools is likely to show even higher error rates for more complex processor
architectures.

\subsection{Dynamic Timing Analysis}
With dynamic timing analysis, the execution time of software is measured 
directly on the hardware. This means that target hardware must be available.
The level of granularity at which these measurements
can be performed varies for different processor architectures. While most
current processor architectures support hardware performance counters, these
counters can only provide a limited level of accuracy and care must be taken in
order to obtain accurate measurements~\cite{Zaparanuks2009}. Moreover, it might
be necessary to modify the observed program by adding instrumentation code for
manipulating the hardware performance counters of the processor. Modifying the
program code can obviously impact the measurements. To perform measurements
with an increased level of accuracy, e.g., up to the level of individual
instruction, additional tools like logic analyzers or processors with dedicated
tracing hardware support are required. 

Probabilistic and statistical timing analyses~\cite{Bernat2002} are variants of
dynamic approaches for execution time estimation. Probabilistic timing analysis
tries to capture the variance of execution times by providing a probability
distribution for the possible execution times of a program. This distribution is
calculated by analyzing the execution time of individual program parts from a
large set of measurements. The complete measurement process can take several
days of observing the system in operation and produce gigabytes of data. Using
this data, the execution time distributions of smaller program parts are
incrementally combined to get the distribution for the complete program. The
limitation of this combination step is that the execution time of individual
program parts is assumed to be independent, which is often not the case in
practice.
More recent approaches for dynamic timing analysis apply statistical methods,
like extreme value theory, to reason about the worst-case execution of a
program without ever observing it~\cite{Lu2012}.  However, this is still an
area of active research without a generally accepted solution.


\subsection{Integration into the Development Process}
In some application domains, e.g, avionics or automotive, worst-case execution
time analysis is standard practice during software development to verify that
real-time requirements are met. Special
software development tools and languages used in the automation domain often
prevent the use of tools from other domains.  Additionally, the real-time
requirements can be slightly different. On the one hand, there are hard
real-time boundaries for the execution time of control applications which are
defined by their cyclic activation. On the other hand, it is acceptable that
these deadlines are missed \emph{sometimes} as long as this does not happen
repeatedly.  This can happen if an isolated event, which is not part of the 
typical execution path, requires some special handling in the code.  Examples
for this are the execution of initialization code during startup or the
creation of status messages.  Since the system design can handle this
application behavior, missing an execution deadline for other reasons, like an
increased execution time due to cache misses, is less critical than in other
application domains which require \emph{guaranteed} upper bounds for the
execution time of software components.  However, WCET estimates reported by
existing analysis tools only consider the \emph{theoretical} worst-case
execution path through a program, which is executed very rarely (or not at all)
for typical control applications.  If the WCET is used to characterize the CPU load
of a control device, this leads to a significant waste of computing resources.

A static analysis technique for average-case execution time would be a viable
alternative. Finding the typical execution path through a program
requires execution of the program or additional manually provided information
about the expected behavior of the program. The latter is currently not done in
any solution. Furthermore, the potential variation between the best-, average-,
and worst-case execution times is an important piece of information which
should be made available as a result of an execution time analysis. Currently,
there is no system or tool which can reason about this variation in a static
way, meaning a way which does not require the execution of the program of
interest.

\section{New Approach: Timing Model Generation for Control Applications}\label{sec:timing}
A major hurdle for estimating the execution time of cyclic real-time control
applications is that, in the current development process, these estimates can
only be made as soon as an industrial control device (controller) is available.
Static timing analysis tools are currently not available for all the
processor variants.  Consequently, the CPU load on the device can only be
determined after the control applications have been developed and deployed to
hardware.  The load is defined as
$$\mathit{load}=\frac{\mathit{application~execution~time}}{\mathit{periodic~activation}} $$
The periodic activation depends on the production process being regulated by the control logic, meaning how often
the application has to check inputs and adjust adjust.
Knowing the expected load early, \ie already during application development is beneficial for two reasons: As
there is a direct correlation between the processing power of a controller and
its cost,  it is very desirable to choose a hardware variant which matches
the application requirements without being overbuilt. Moreover, with a foreseen
increase of distributed or parallel execution of applications, information
about the execution time of individual program parts helps in tailoring the
control application to its execution environment, \eg through efficient
partitioning of an application on a controller with a multi-core processor. To
achieve this, execution time estimation must be an integral part of the
software development process.

\subsection{Timing Model Generation Flow}

The intention of the proposed approach is to replace the currently
used abstract, but complete models of CPU timing behavior with a model which
only covers those instruction sequences which are actually used by programs in
the targeted domain. This approach can significantly reduce the cost for
developing such models and also makes them applicable in domains with less
strict requirements.  In contrast to similar work presented
in~\cite{Altenbernd2011}, the proposed technique is not intended to generate a
general purpose timing model for arbitrary applications.  A similar idea for
reusing WCET estimates has already been presented in~\cite{ReusableWCET}, but
the approach presented in this work aims at a reuse at a more fine-grained
level.

To accurately estimate the best-, average- and worst-case execution time of a
software component without requiring its execution to be observed, a timing
model for the processor executing the software must be available.  The
execution time of a sequence of machine instructions greatly depends on how
these instructions move through the pipeline and the functional units of the
processor and how their execution can be interleaved. The execution time of
individual instructions, e.g., from the processor manual or from repeatedly
measuring the execution of the same instruction type, can only serve as a
starting point for estimating the execution time of a complete program. On the
other hand, the way machine code for control application is generated, e.g,
when using ABB's family of Control Builder tools or similar model-based tools
allows the reasonable assumption that certain sequences or patterns of
instructions occur repeatedly.  Thus, in contrast to known static analysis methods,
a general model of the processor pipeline
might not be needed to obtain reasonably accurate performance estimates for
control applications.

This insight motivates using their domain-specific properties to characterize the
execution time of control applications based on  recurring code sequences.
The basic idea is to
\begin{enumerate}
\item automatically extract recurring code sequences from a set
of training applications and
\item to use automatic test data generation techniques to determine their on-target execution time. 
\end{enumerate}
Thereby, a timing model for the CPUs used in industrial control devices 
can be generated with acceptable effort. Still, this model provides sufficient timing
accuracy. The timing model, i.e., the contained timing information about recurring
code sequences, then is used to determine the execution time of control
applications with a similar structure. Moreover, the model is also applicable
to applications that are created using the same set of tools.
This approach is thus an alternative to the current practice of manually
developing timing models, e.g., as used in commercial tools for static timing
analysis.  

\begin{figure}
\centering
\includegraphics[height=.8\textwidth, page=1]{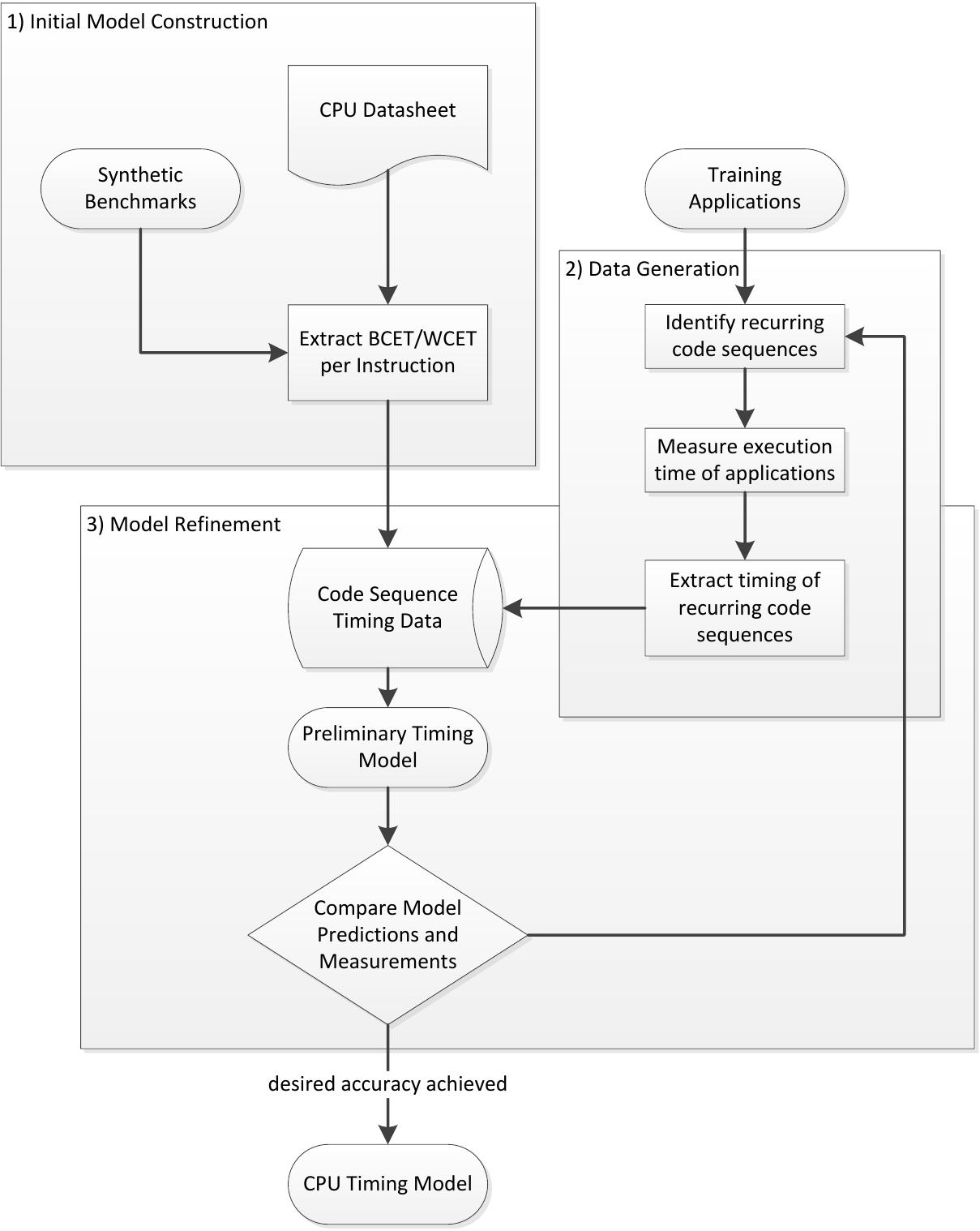}
\captionof{figure}{Timing Model Construction}
\label{fig:model_construction}
\end{figure}

The proposed timing model generation work flow is shown in
Figure~\ref{fig:model_construction}. An initial timing model of a CPU is
constructed by considering single instructions only in step~1. This initial
model should at least contain best-case and an worst-case execution time
estimates for every instruction type of the CPU for the target controller. This
information can be extracted from the CPU data sheet or by using synthetic
benchmarks to measure the execution time of a single instruction type. This
step has to be repeated for every new CPU variant used in an industrial
controller. The resulting instruction-level estimates could already be used to
automatically determine coarse-grained performance estimates for
every control application as long as the occurrence of individual machine
instructions can be counted. 

Subsequent to the creation of the baseline model, which only considers individual
instructions, data to create the timing model of the CPU is created using a set of control
applications to train the model, as shown by step~2 of
Figure~\ref{fig:model_construction}. This refinement is done by extending the timing model to
longer sequences of machine instructions.  The training applications are
decomposed into smaller code pieces which are searched for recurring code
sequences. For these recurring sequences, more precise estimates are obtained
by performing detailed measurements of the respective program parts. To allow for
estimating variations in the execution time of a program, the implementation
must be able to track the execution time triple of each code sequence, meaning
the best-case, the average-case and the worst-case execution times.  Ideally,
the model should not only consider the execution time of code sequences in
isolation, but also their interleaving and interdependence. That is, if the
execution time of a code sequence \emph{A} is influenced by the fact that code
sequence \emph{B} is executed in advance, this information is also considered
by the model.  After the timing model has been created automatically, standard static
analysis techniques can be used to determine the possible execution of an
arbitrary control application.  The structure of the control application has to
be decomposed into code sequences for which the timing model can provide
execution time estimates. If no information for longer code sequences is
available in the timing model, it might be necessary to decompose the program
up to the level of individual machine instructions. The latter case is always
possible and can make use of the information from the baseline model.  Using
the timing model created by the proposed approach, the execution time of newly
developed applications can be performed through a bottom-up accumulation of
execution times of recurring code sequences, e.g., based on the control flow
graph of the program.  The process of finding recurring code sequences and
characterizing their execution time will be described in the following
paragraphs. The model refinement continues until the predictions of the generated model fulfill
the desired timing accuracy (step~3 in Figure~\ref{fig:model_construction}),
which can be checked by comparing the predictions of the model to end-to-end
execution time measurements of the training applications.

\subsection{Finding Recurring Code Sequences}
\label{recurring_code}
The fundamental requirement for the proposed timing model generation is that
recurring code patterns exist in the application code and that these patterns
can be detected automatically. Code clone detection techniques are a natural
choice to detect these patterns in the machine code.  However, most code clone
detection techniques work at the source code level, although there are
publications dealing specifically with binary code~\cite{Saebjornsen2009}.  The
simplest form of searching for recurring code sequences is by looking for
verbatim copies.  For source code, this would mean searching for identical
character sequences in the source code, but discarding white space.  The
binary-level equivalent for this approach would be to search for sequences in
the machine code which are identical down to the last bit.

As two pieces of code with the same origin and identical functionality do not
have to be completely isomorphic, e.g., due to some renaming of variables in the
source code or using different registers for the same machine code operations,
code clone detection techniques usually apply different forms of normalization
to the analyzed program.  For code clone detection at the machine code level,
normalization techniques include discarding the order of instructions,
abstracting the opcodes into certain classes, or translating the arguments of
machine code operations into a symbolic form.  These normalization techniques
are applied to certain code areas, e.g., instruction sequences of a fixed
length.  The result can be a vector of attributes, for instance a vector
containing number of instructions with a given opcode, or a hash value of the
normalized instruction sequence.  This normalized representation can be used to
efficiently compare code areas to detect similar code areas and thus,
(potential) code clones.

\begin{figure}
\centering
\includegraphics[height=.35\textwidth]{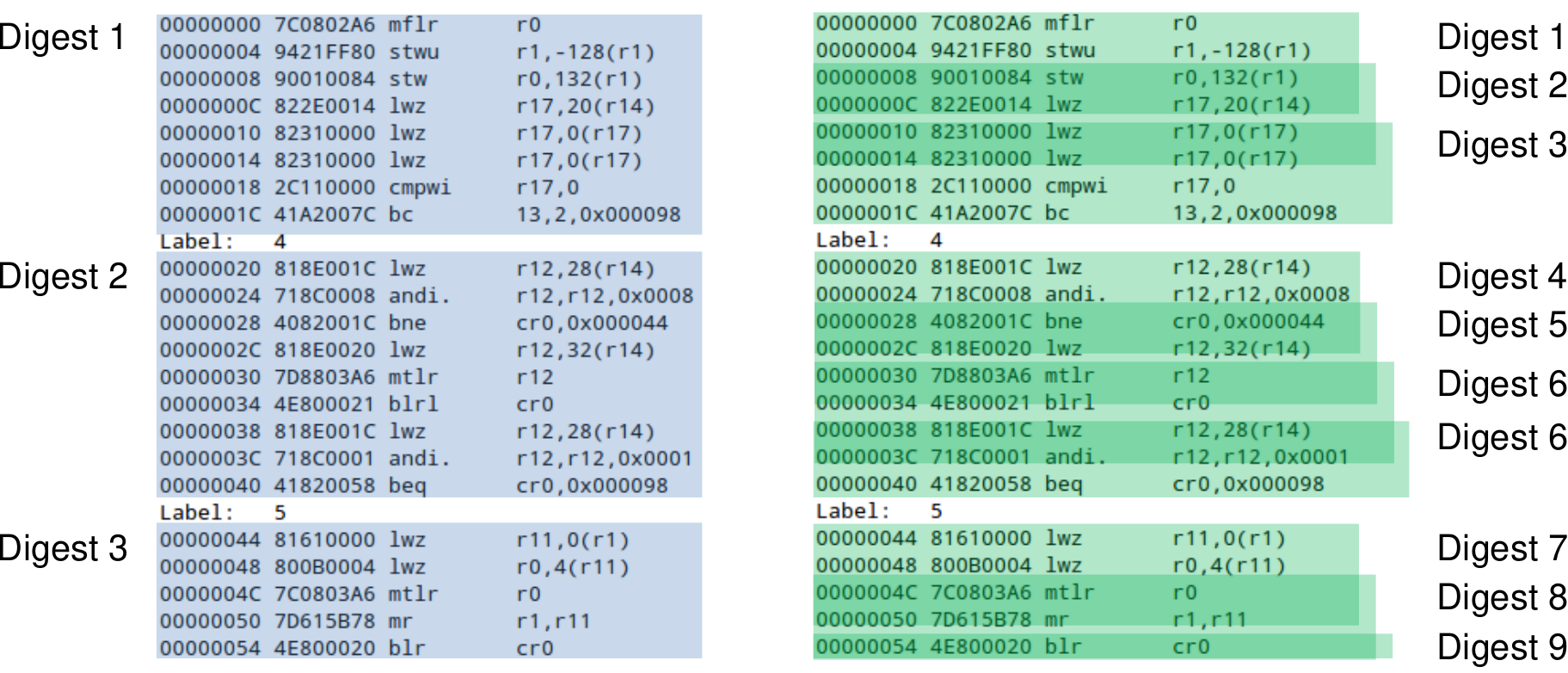}
\captionof{figure}{Basic Block and Window-Based Digest Computation}
\label{fig:digests}
\end{figure}

For the purpose of characterizing the execution time of recurring code
sequences there are several constraints for applying known normalization techniques:
The order of machine instructions should not be discarded as the
execution order has a significant impact on execution time. The same holds for
operands of an instruction, but to a lesser extent. 
To simplify the detection
of recurring code sequences, the detection algorithm
focuses on sequences of instruction opcodes only.  Thus, an MD5 hash for the
opcodes within basic blocks is computed and used as normalized characterization
of the application code. For a given sequence of machine instructions,
including their address, input registers and intermediate constants, only the
opcode of the instruction is used for computing the MD5 digest of the sequence.
Basic blocks can be further decomposed when searching for recurring sequences,
but the sequences used for clone detection cannot span multiple basic blocks.
Therefore each considered sequence can contain at most one branch instruction.

Different granularities for the computation of digests are illustrated in
Figure~\ref{fig:digests}.  The machine code shown was generated by
ABB's engineering tool Compact Control Builder, which is used to develop control
applications.  For the most coarse-grained variant on the left side, all
instructions of a basic block are used directly.  This means that the value of
Digest 1 is computed by applying the MD5 hash function to the opcode
sequence~(\texttt{mflr}, \texttt{stwu}, \texttt{stw}, \texttt{lwz},
\texttt{lwz}, \texttt{lwz}, \texttt{cmpwi}, \texttt{bc}).  For all other basic
blocks, which are marked with a label in the code, the digest is computed in
the same way.

For a more fine-grained characterization, digests can be calculated using a
sliding window approach which splits up the basic blocks in the machine code.
To generate instruction sequences from a basic block, sequences are determined
by moving a window of fixed size across the basic block using a fixed stride.
This is illustrated in the right-hand part of Figure~\ref{fig:digests} using a
window size of 4 instructions and a stride of 2 instructions. For Digest~1, the
hash value of the opcode sequence (\texttt{mflr}, \texttt{stwu}, \texttt{stw},
\texttt{lwz}) is computed.  The stride value can only be less than or equal to
the window size, because otherwise not all instructions from the basic block
would be included. Choosing the stride value smaller than the window size makes
it more likely to capture recurring sequences. At the end of a basic block,
the window size is pruned if it would otherwise move across the end of the
basic block. For the last window of this example, the window size is pruned and
thus Digest 9 is computed only from the single opcode \texttt{blr}.

After the training applications for timing model generation have been
partitioned into digests, the digest values can be used to detect recurring
code sequences. The underlying assumption is that sequences with identical
digests will have identical execution times. To reduce the measurement effort
needed to characterize the execution time of the recurring sequences, the
number of sequences contained in the timing model should be as small as
possible. On the other hand, the model should contain enough instruction
sequences so it is rarely necessary to fall back to the single instruction
baseline model when characterizing the execution time of a newly developed
control application.  This trade-off is still under investigation. Preliminary
results on how to detect the recurring code sequences for an accurate and
minimal timing model are presented in section~\ref{experiments}.

\subsection{Timing Characterization}
Determining the execution time of recurring code sequences poses several
challenges.  First of all, it must be possible to measure the execution time of
relatively short instruction sequences. On most modern processor architectures,
this is only possible by adding instrumentation code.  When characterizing the
execution time of recurring code sequences, there is obviously a trade-off
between the effort needed to perform the measurements and the accuracy of the
characterization.  In particular the overhead added to measurements by adding
instrumentation code has to be considered.  Based on our initial evaluation,
the most suitable framework for implementing this part of our proposed work flow is the Dyninst binary
instrumentation infrastructure~\cite{Dyninst}, as it provides static
instrumentation capabilities which do not add virtualization overhead to the
measurements. A prototype tool based on Dyninst is currently being implemented.

Another challenge for the timing characterization is observing all relevant
execution times. This will be tackled by extensive testing during model generation.
One option is to use the existing test cases for the training
applications from which the timing model is generated. 
As the main purpose of such test cases is not the generation of a CPU timing model,
additional tests will be necessary.  It is planned to apply automatic test case
generation techniques. Thus, approaches like random or concolic
testing~\cite{Godefroid:2005:DDA:1065010.1065036,
Pacheco:2007:RFR:1297846.1297902} will be used to generate input data for the measurements
in the final implementation of the proposed approach.  The recently introduced
concept of micro execution~\cite{Godefroid2013} to test small portions of
binary code with arbitrary inputs might also be applicable.

The important difference between the proposed work flow and existing
measurement-based timing analysis tools is that timing measurements have to be
performed only once per target processor and not once per application. The
effort for setting up a test environment and measurement facilities will
therefore only be needed once.
The execution time measurements are formalized by applying sophisticated test case generation techniques.
Furthermore, it is not necessary that the user
of the resulting timing model has access to the target hardware to generate
timing estimates for a newly developed application.
The second important property is the integration of
average-case execution times for the recurring code sequences into the model.
This allows reasoning about the average load of the targeted device without
running the software on the actual hardware. To achieve this,
the model must store the best-case, the average-case and the worst-case
execution time for each recurring code sequence.

\section{Execution Time Estimation with User Feedback}\label{sec:user}
In order to present the information about the execution time to the developer
and to allow a fine-grained consideration of the program structure in the
execution time analysis, a tight integration into the software development
environment is crucial.  An approach for integrating WCET estimates into a code
editor to give the software developer feedback about the execution time of a
program has been presented in~\cite{Harmon2012}. By contrast, the approach
presented in this paper makes the possible variation in the execution time of a
software component explicit and thus enables the developer to make more
accurate decisions whether all real-time requirements of the application can be
met under the expected conditions.

After a timing model has been created for a target processor using a set of
training applications, the resulting model can be used to estimate the
execution time of newly developed applications. Thus, it could serve as a
replacement for the micro-architectural analysis of existing analysis tool chains
described in section~\ref{sa_timing}.  However, our approach might still lead to overly
pessimistic execution time estimates due to incorrectly approximated program
control flow. Nonetheless, the previously described approach for an automatic
timing model generation overcomes the need for additional measurements and access to hardware
during software development, while the effort for developing a timing model is reduced.

\begin{figure}
\centering
\begin{minipage}{.5\textwidth}
  \centering
  \includegraphics[height=.7\textwidth]{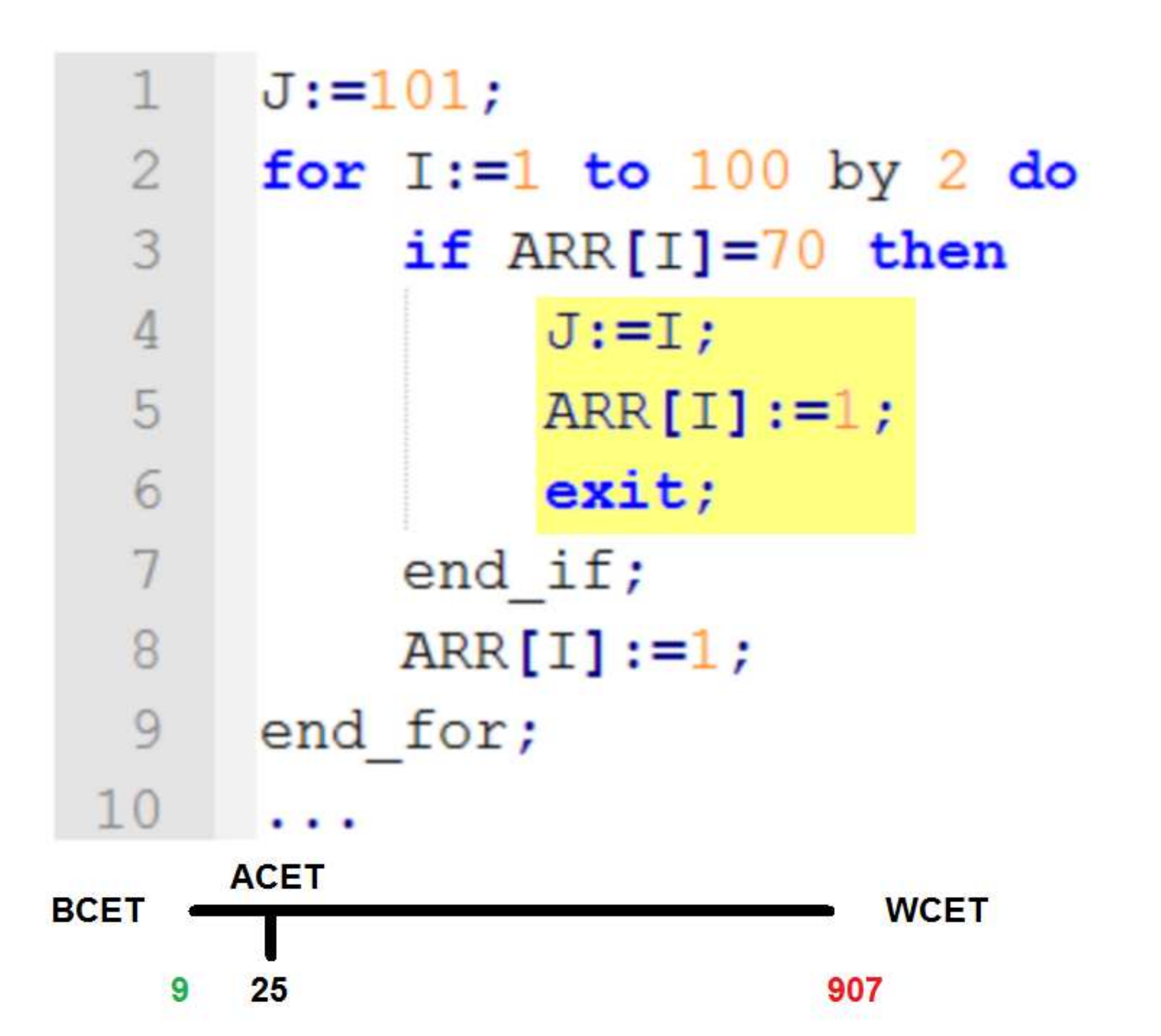}
  \captionof{figure}{Typical Execution Marked}
  \label{fig:sample_code_highlight}
\end{minipage}%
\begin{minipage}{.5\textwidth}
  \centering
  \includegraphics[height=.7\textwidth, page=4]{CFG.pdf}
  \captionof{figure}{Typical Case Execution Time Estimation}
  \label{fig:CFG4}
\end{minipage}
\end{figure}

To facilitate the use of the timing model, we propose a 3-valued representation of
the potential execution times. This directly presents the application's ACET,
BCET and WCET to the user. Thereby, the user is much better informed about
the application's potential behavior already in the development phase and additionally
gets direct feedback on application changes. A mock-up of how this could look like
is shown at the bottom of Figure~\ref{fig:sample_code_highlight}. 
In addition to giving feedback to the user, the average-case execution time estimate can
be significantly improved by only considering parts of the application which are included in a typical
execution. Therefore, the user has to provide the required 
insights about the expected or \emph{typical} program path---which often can be done due
to the user's experience and knowledge of the application. In combination with
the previously described ACET estimates for recurring code sequences \emph{and} structural program
information, a more accurate approximation of the typical case can be achieved.
The reason for this is that execution time outliers are less likely to impact
the result of the average-case estimate. The standard way of simply averaging
the executing time of observed program runs to get an ACET estimate would still
include such outliers, but the proposed approach allows excluding such atypical
executions.  Finally, the 3-valued representation of the execution times
still makes the worst case explicit to the user and thus, no information is
lost compared to a standard WCET analysis.

The example program from section~\ref{sa_timing} will be used to illustrate the application of user
feedback in estimating execution times. If we assume that the programmer of the
code shown in Figure~\ref{fig:sample_code_highlight} knows that conditional
expression in the loop typically evaluates to \textbf{true} immediately, she
can mark the respective code area as being part of the typical execution.  This
is illustrated in Figure~\ref{fig:sample_code_highlight} by shading the
respective code area. A similar tagging functionality could be easily added to
any code editor.  While this is an artificial example, similar patterns can
often be found in control applications, \eg for initialization code.  By
automatically translating this information into constraints for the global
bound calculation analysis step, it can be used to restrict the program paths
through the CFG.  Thus, the program path determined by the global analysis for
the ACET step no longer considers the worst-case path, but the typical path.
The resulting program path for the example is shown in Figure~\ref{fig:CFG4}.
The edges drawn as dotted lines and the frequency annotations next to them describe
the typical path. Consequently, each of the basic blocks along this path is executed
exactly once. Using this information and the execution time of each basic block,
which is also annotated to the CFG in Figure~\ref{fig:CFG4}, an ACET
of 25 cycles can be derived. The respective ACET formula is shown in the lower portion of Figure~\ref{fig:CFG4}.
Even if there is only small difference between the average-case and the worst-case execution time
of the basic blocks, considering information about the typical program path can
still reduce the overestimation of static timing analysis. While existing WCET
analysis tools already support similar path constraints, the underlying timing
model is always a pure worst-case analysis.


The combination of the ACET estimates and the user-provided information about the typical program path
allows for characterizing the typical execution time of programs
more accurately than with existing solutions. Thus, it goes beyond classical BCET and WCET analysis.
Since the results of the WCET and BCET analysis
are still represented, the information is still available to the user of the
timing analysis. Adding average-case information allows to additionally reason
about the load of a computer system by considering typical execution paths only.
The proposed approach depends, however, on the user's correct knowledge on the typical
execution path. Thus, we are planning to case-study the quality of user input and
the impact of wrong assumptions, \eg mistakenly unmarked typical code areas or wrongly marked
code areas, on calculated ACET estimates with real industrial applications and users.

The 3-valued execution time estimation can be applied at different levels of
granularity, \eg complete applications, components, or individual source
code lines. We envision an integration of this representation into the
integrated development environment for the developers of control applications.
This is illustrated in the lower portion of
Figure~\ref{fig:sample_code_highlight}.  By integrating this representation
into the application development environment, the contribution of individual
program parts is directly presented to the application developer. Thereby the
3-valued representation makes the possible variation of execution times
explicit to the developer. In addition, program parts with a high contribution
to the execution time are highlighted directly if the 3-valued execution time
estimate is represented at the level of basic blocks or individual code lines.
To adapt the 3-valued execution time estimate, the developer can mark program
areas as being not part of a typical execution.  This will exclude the
respective program parts from the average-case execution time estimate.
Alternatively, the offending program parts can be optimized manually by the
developer and the 3-valued estimate can instantly provide information about the
impact of the code changes on the execution time.

In the context of industrial control applications, code portions which belong
to an atypical execution path can often be identified automatically. The result
of a certain firmware function directly relates to whether the controller is in a
typical state or not. For instance in most executions of a control application,
the firmware function which checks whether the device just experienced a warm
restart will return \textbf{false}. Thus, when the respective function is used
in a conditional expression, this can be directly translated into a constraint
for the average-case analysis. When performing a worst-case analysis, these constraints
should of course not be used as they might lead to an underestimation of the WCET.

Essentially, marking certain code areas as the typical execution path excludes
alternative paths from the WCET analysis and thus approximates the expected
behavior more accurately.  Depending on the application for which the execution
time is estimated, the user-provided information could also be used to
\emph{tag} code areas for other scenario-based analyses. This allows for a more
accurate estimation of the execution time under certain preconditions.  The
proposed approach could therefore also be extended for analyzing the execution
time of a program under specific operating conditions.

\section{Experimental Results}\label{sec:results}
\label{experiments}

This section presents preliminary results for the proposed timing model
generation work flow.  The prototyping was carried out using the
machine code output facilities of ABB's Compact Control Builder development
tool targeting the ABB AC~800M controller family. The basic block and
window-based hash computation described in Section~\ref{recurring_code} were
applied to three real-world Compact Control Builder projects to evaluate the
amount of code sharing across different applications.

Each project contained multiple control applications which  shared some
code through common libraries. However, these libraries are always reused
as source code, not as compiled machine code.
Binary-level code sharing was analyzed by first calculating
all possible digests for a fixed calculation technique for two of the three
projects.  All of the possible digests were stored in a database. Then all
possible digests of the remaining project were computed and compared to the
information stored in the database. The outcome of this experiment is shown in
Table~\ref{table:granularity} for the basic block digest computation and
window-based digest computation using a window size of 8 and 16
instructions.  The size of each project is given in terms of instructions in
the machine code after compilation. If certain modules were reused across
different applications in the same project, the respective machine instructions
were only counted once.  The detection of recurring code patterns across the
different projects could be executed in a few minutes on a standard laptop
computer.  Since the analysis tools is a very early prototype, no detailed
performance measurements were conducted.

\begin{table}
\begin{tabular}{|l|c|p{1.6cm}|p{1.6cm}|p{1.6cm}|p{1.6cm}|p{1.6cm}|p{1.6cm}|}
\hline
		&	& \multicolumn{2}{|c|}{\textbf{Basic Blocks}}	& \multicolumn{2}{|c|}{\textbf{Window=16, Stride=8}}	& \multicolumn{2}{|c|}{\textbf{Window=8, Stride=4}} \\
\hline
\textbf{Project ID} & Size & Recurring Patterns & Instruction Coverage & Recurring Patterns & Instruction Coverage & Recurring Patterns & Instruction Coverage \\
\hline
Project 1	& 53028		& 3024	& 54.13\%	& 5902	& 82.50\%	& 12634		& 94.34\% \\
Project 2	& 545328	& 14593	& 65.16\%	& 64058	& 88.22\%	& 134097	& 97.55\% \\
Project 3	& 119546	& 4309	& 62.51\%	& 14213	& 87.35\%	& 29634		& 97.37\% \\
\hline
\end{tabular}
\caption{Clone Detection Granularity}
\label{table:granularity}
\end{table}

The number of recurring code sequences was calculated by counting the number of
hash digests a project has in common with the two remaining projects.  So the
numbers given in each line of the table show how much recurring code patterns
where found in the respective project when the other projects where used as
training data.  Besides counting the number of recurring code sequences for the
respective clone detection granularity, the second result column for each
variant contains the percentage of machine instructions covered by these
sequences. This value is calculated by dividing the number of machine
instructions within a project which are included in recurring code sequences by
the total number of instructions in the project.  The rationale behind this estimation is that it determines the
maximal achievable coverage if all recurring code sequences of the respective
granularity would be part of the timing model.

The instruction coverage is relevant since it is not only an indicator of how
much code sharing between different projects could be detected, but also what
amount of code was not covered by any of the recurring code sequences. For the
proposed automatic timing model generation, the number of recurring code
patterns covered by the model should be minimal, while the instruction coverage
on the training applications should be maximized.  What can be seen from the
results is that although there is a significant amount of code sharing, even
when only considering basic blocks, the amount of sharing is insufficient to
derive an execution time model from basic block reuse. 

To further investigate the amount of code sharing inside these projects,
further analyses were conducted using a window-based digest computation with a
stride parameter of 1. Essentially this means that all possible combinations
for the respective window size which actually occur in the training data where
tested.  The intention of this experiment was to investigate whether the
number of recurring code patterns are inversely proportional to the window size
or whether the number stabilizes at some point. The results for the second
experiment are shown in Table~\ref{table:windowsize}.  Choosing the minimal value for the
stride parameter achieves higher instruction coverage while creating a larger
number of recurring code sequences. While the number of recurring code sequences
continues to grow, a window size of eight instructions creates an almost complete
instruction coverage. The optimal window size thus should be in the order of ten instructions.

What has not been investigated so far is which of the recurring code patterns
are actually needed to create a timing model which can accurately describe the
execution time of control applications.  With appropriate heuristics it should
be possible to select a small percentage of these recurring code patterns and
still reach a good instruction coverage. Covering a large percentage of the
instructions in a control application is necessary to derive a model with good
timing accuracy, but the more instruction sequences the model contains the more measurements
have to be taken to create the model.

\begin{table}
\begin{tabular}{|l|p{1.8cm}|p{1.8cm}|p{1.8cm}|p{1.8cm}|p{1.8cm}|p{1.8cm}|}
\hline
			& \multicolumn{2}{|c|}{\textbf{Window=16, Stride=1}}	& \multicolumn{2}{|c|}{\textbf{Window=8, Stride=1}}	& \multicolumn{2}{|c|}{\textbf{Window=4, Stride=1}} \\
\hline
\textbf{Project ID}	& Recurring Patterns & Instruction Coverage & Recurring Patterns & Instruction Coverage & Recurring Patterns & Instruction Coverage \\
\hline
Project 1	& 46278	& 93.27\%	& 50344	& 98.68\%	& 52678		& 99.89\% \\
Project 2	& 503423	& 97.20\%	& 533533	& 99.67\%	& 544413	& 99.99\% \\
Project 3	& 108878	& 96.52\%	& 116403	& 99.55\%	& 119228		& 99.99\% \\
\hline
\end{tabular}
\caption{Clone Detection Window Size}
\label{table:windowsize}
\end{table}

\section{Conclusion and Future Work}\label{sec:conc}
This paper presented techniques for automatically creating a timing model for a
given  processor and for the static estimation of average-case software
execution times based on user feedback. 
These concepts were originally developed in the context of
industrial control applications, but they likely are also applicable in other fields
of software development.  The assumption that control applications contain a
significant amount of recurring code sequences has been verified by initial
experiments based on real industrial applications. The results also show that a certain level of
decomposition must be used to derive generic models for software execution
times on a given processor, e.g., using basic blocks as the
level of granularity for timing model generation is insufficient for reaching
the desired coverage. Thus, only focusing on direct code reuse, e.g, through
software library types which are reused across many applications, will not be
sufficient to generate an accurate timing model.  Characterizing instruction
sequences using a sliding window approach is more appropriate, although it
potentially creates a lot of overlapping code patterns. It should be possible
to overcome this issue if the timing model generation considers instruction
coverage when generating sliding windows. Code sequences generated from the
sliding window approach should only be integrated to the model if instruction
coverage is improved.

The recurring code sequences, which are the basis of an automatically generated timing model,
are ideally in the order of ten instructions. This fact implies
that performance measurements should ideally be possible at the same level of
granularity. Alternatively, the execution time of shorter code sequences must
be extracted from measurements for longer sequences. Since the former is
unlikely for a complex processor design, the latter will be investigated in
more detail in future work. We plan to validate the accuracy of the resulting model
in an industrial case study based on our promising initial results.

Another interesting aspect is the relation between the recurring code sequences
considered by the timing model and the provided timing accuracy. Minimizing the
number of code sequences necessary to achieve a certain level of code coverage
on the training applications might not be the optimal solution. Instead, the
impact on timing accuracy should already be considered when selecting recurring
instruction sequences to  be part of the timing model. Implementing this in the
proposed model generation flow will also be addressed in future work.

\nocite{*}
\bibliographystyle{eptcs}
\bibliography{bibliography}
\end{document}